\documentclass[fleqn,usenatbib]{mnras}
\usepackage[dvips]{color}
\usepackage[dvipsnames]{xcolor}
\usepackage{ulem}
\usepackage{longtable}
\usepackage{url}
\usepackage{epstopdf}
\usepackage{graphics}
\usepackage{graphicx}
\usepackage{caption}
\usepackage{times}
\usepackage{lineno}
\usepackage{xspace}
\usepackage{amssymb,amsmath}
\usepackage{longtable}
\definecolor{orcidlogocol}{HTML}{A6CE39}

\usepackage{graphicx}
\usepackage{lscape}
\usepackage{xcolor}
\usepackage{lineno}

\def\arcsec{\hbox{$^{\prime\prime}$}}

\def\apj{ApJ}

\def\aap{A\&A}
\def\mnras{MNRAS}
\def\apjs{ApJS}

\def\nat{Nature}
\def\apjl{ApJLett}

\newcommand{\GRB}{{GRB\, 191221B\,}\xspace}

\begin{document}
\title[GRB191221B]
{Spectropolarimetry and photometry of the early afterglow of the gamma-ray burst  \GRB \thanks{Based on observations made with the Southern African Large Telescope (SALT) and the MeerKAT radio telescope array.
}}
\author[Buckley et al.]{D. A. H. Buckley$^{1,2, \href{https://orcid.org/0000-0002-7004-9956}{\includegraphics[scale=0.7]{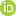}}}$\thanks{E-mail: dibnob@saao.ac.za}, 
S. Bagnulo$^{3}$,
R. J. Britto$^{4}$,
J. Mao$^{5,6,7}$, 
D. A. Kann$^{8,
\href{https://orcid.org/0000-0003-2902-3583}{\includegraphics[scale=0.75]{figures/ORCIDiD_icon16x16.png}}}$,
J. Cooper$^{4}$,
\newauthor 
V. Lipunov$^{9,10}$,
D. M.  Hewitt$^{1,2}$,
S. Razzaque$^{11}$, 
N. P. M. Kuin$^{12,\href{https://orcid.org/0000-0003-4650-4186}{\includegraphics[scale=0.75]{figures/ORCIDiD_icon16x16.png}}}$, 
I. M. Monageng$^{1,2}$,
\newauthor 
S. Covino$^{13}$, 
P. Jakobsson$^{14}$,
A. J. van der Horst$^{15,16}$,
K. Wiersema$^{17,18}$,
M. B\"ottcher$^{19}$,  
\newauthor 
S. Campana$^{13}$, 
V. D'Elia$^{20,21}$,
E. S. Gorbovskoy$^{9}$,
I. Gorbunov$^{9}$, 
D. N. Groenewald$^{1,22}$, 
\newauthor 
D. H. Hartmann$^{23,\href{https://orcid.org/0000-0002-8028-0991}{\includegraphics[scale=0.75]{figures/ORCIDiD_icon16x16.png}}}$,
V. G.  Kornilov$^{9,10}$, 
C. G. Mundell$^{24}$, 
 R.  Podesta$^{25,26}$, 
J. K. Thomas$^{1}$, 
\newauthor 
N. Tyurina$^{9}$, 
D. Vlasenko$^{9,10}$, 
B. van Soelen$^{4}$ and 
D. Xu$^{27}$\\
\\
$^1$ South African Astronomical Observatory, PO Box 9, Observatory Road, Observatory 7935, Cape Town, South Africa \\
$^2$ Department of Astronomy, University of Cape Town, Private Bag X3, Rondebosch 7701, South Africa\\
$^3$ Armagh Observatory \& Planetarium, College Hill, Armagh, Northern Ireland, BT61 9DG, UK\\
$^4$ Department of Physics, University of the Free State, PO Box 339, Bloemfontein, 9300, South Africa\\
$^{5}$ Yunnan Observatories, Chinese Academy of Sciences, 650011 Kunming, Yunnan Province, China\\
$^{6}$ Center for Astronomical Mega-Science, Chinese Academy of Sciences, 20A Datun Road, Chaoyang District, 100012 Beijing, China\\
$^{7}$ Key Laboratory for the Structure and Evolution of Celestial Objects, Chinese Academy of Sciences, 650011 Kunming, China\\
$^{8}$ Instituto de Astrof\'isica de Andaluc\'ia (IAA-CSIC), Glorieta de la Astronom\'ia s/n, 18008 Granada, Spain\\
$^9$ M. V. Lomonosov Moscow State University, SAI, Physics Department, 13 Univeristetskij pr-t, Moscow 119991, Russia\\
$^{10}$ M.V.Lomonosov Moscow State University, Physics Department, Leninskie gory, GSP-1, Moscow 119991, Russia\\
$^{11}$ Centre for Astro-Particle Physics (CAPP) and Department of Physics, University of Johannesburg, PO Box 524, Auckland Park 2006, South Africa\\
$^{12}$ Mullard Space Science Laboratory, Dept. of Space and Climate Sciences, University College London, Holmbury St Mary, Dorking, RH5 6NT, UK\\
$^{13}$ Brera Astronomical Observatory, via Bianchi 46, I-23807, Merate(LC), Italy\\
$^{14}$ Centre for Astrophysics and Cosmology, Science Institute, University of Iceland, Dunhagi 5, 107, Reykjavík, Iceland\\
$^{15}$ Department of Physics, The George Washington University, 725 21st St. NW, Washington, DC 20052, USA\\
$^{16}$ Astronomy, Physics and Statistics Institute of Sciences (APSIS), The George Washington University, Washington, DC 20052, USA\\
$^{17}$ Department of Physics, Gibbet Hill Road, University of Warwick, Coventry, CV4 7AL, UK\\
$^{18}$ Department of Physics and Astronomy, University of Leicester, LE1 7RH, UK\\
$^{19}$ Centre for Space Research, North West University, Potchefstroom, South Africa \\
$^{20}$ ASI-Space Science Data Center, via del Politecnico snc, 00133 Rome, Italy\\
$^{21}$ INAF - Osservatorio Astronomico di Roma, via Frascati 33, 00040 Monte Porzio Catone, Italy\\
$^{22}$ Southern African Large Telescope Foundation, PO Box 9, Observatory Road, Observatory 7935, Cape Town, South Africa \\
$^{23}$ Department of Physics and Astronomy
Clemson University, Clemson, SC 29634-0978, USA \\
$^{24}$ Department of Physics, University of Bath, Claverton Down, Bath BA2 7AY, UK\\
$^{25}$ Observatorio Astronomico Felix Aguilar(OAFA), Avda Benavides s/n, Rivadavia, El Leonsito, Argentina\\
$^{26}$ San Juan National University, Casilla de Correo 49, 5400 San Juan, Argentina\\
$^{27}$ CAS Key Laboratory of Space Astronomy and Technology, National Astronomical Observatories, Chinese Academy of Sciences, Beijing 100101, China}


\pubyear{2021}

\date{Accepted XXX. Received YYY; in original form ZZZ}

\label{firstpage}

\pagerange{\pageref{firstpage}--\pageref{lastpage}} 

\maketitle

\newpage\clearpage

\begin{abstract}
We report on results of spectropolarimetry of the afterglow of the long gamma-ray burst GRB 191221B, obtained with SALT/RSS and VLT/FORS2, as well as photometry from two telescopes in the MASTER Global Robotic Network, at the MASTER-SAAO (South Africa) and MASTER-OAFA (Argentina) stations. Prompt optical emission was detected by MASTER-SAAO 38 s after the alert, which dimmed from a magnitude (white-light) of $\sim10$ to 16.2 mag over a period of $\sim$10 ks, followed by a plateau phase lasting $\sim$10 ks and then a decline to $\sim$18 mag after 80 ks. The light curve shows complex structure, with four or five distinct breaks in the power-law decline rate. SALT/RSS linear 
spectropolarimetry of the afterglow
began $\sim2.9$ h after the burst, during the early part of the plateau phase of the light curve. Absorption lines seen at $\sim6010$ \AA~ and 5490 \AA~ are identified with the Mg II 2799 \AA~ line from the host galaxy at $z=1.15$ and an intervening system located at $z=0.96$. The mean linear polarisation measured over $3400-8000$ \AA~ was $\sim$1.5\% and the mean equatorial position angle ($\theta$) $\sim$65$^{\circ}$. VLT/FORS2 
spectropolarimetry was obtained $\sim10$ h postburst, during a period of slow decline ($\alpha=-0.44$), and the polarisation was measured to be $p=1.2$\% and $\theta=60^{\circ}$.
Two observations with the MeerKAT radio telescope, taken 30 and 444 days after the GRB trigger, detected radio emission from the host galaxy only.
We interpret the light curve and polarisation of this long GRB in terms of a slow-cooling forward-shock. 
\end{abstract}

\begin{keywords}
High energy astrophysics; Gamma-ray bursts; Magnetic fields; Polarimetry; Shocks; Jets
\end{keywords}

\section{Introduction}
Gamma-ray bursts (GRBs) are fast, high-energy transient phenomena which, during the sub-second to few hundred
seconds duration of the event, are the most luminous sources of $\gamma$-rays in the Universe, with a typical energy release of $\sim$10$^{51}$ ergs.
GRBs are the result of the collapse of massive, highly evolved stars, or the merger of compact objects,
with a significant number, particularly the so-called ``long-soft'' GRBs, linked to core-collapse supernovae (for a review, see \citealt{Cano2017}). Accretion onto a resulting compact object, like a black hole or neutron star, produces powerful ultra-relativistic jets which, through dissipation processes like shocks or magnetic reconnection, produce prompt $\gamma$-ray emission (for reviews of GRBs and GRB physics, see e.g. \citealt{Gehrels2009}, \citealt{2013FrPhy...8..661G}, \citealt{Gao2013}, \citealt{Wang2015}, \citealt{Kumar2015}). 

The resulting rapidly expanding ejecta of a GRB, after the prompt emission phase, collides with the surrounding medium, producing long-lasting emission called an afterglow, detected across the 
whole electromagnetic spectrum (e.g., \citealt{Piran1999, Meszaros2002, Piran2004}). At the onset of the collision-driven  afterglow, shocks are formed, one forward-propagating into the external medium, while another, shorter-lived
reverse shock propagates backward into the jet \citep{SariPiran1999, Kobayashi2000}. The interaction between the ejecta and the surrounding medium may be quantified by 
several micro-physical parameters, such as the degree of the ejecta's magnetisation, $\sigma_B$. This is the ratio of magnetic to kinetic energy and in the matter-dominated regime model for a standard fireball, $\sigma_B$ $<$ 1, and therefore shocks are plasma-dominated \citep{ReesMeszaros1994,Gomboc2008}.
With increasing $\sigma_B$ the magnetic energy becomes significant, 
and the reverse shock develops until it reaches a maximum at $\sigma_B$ $\sim$ 0.1, whereupon it weakens and is suppressed for $\sigma_B$ $\geq$1 (\citealt{Giannios2008} and references therein). For a highly magnetised outflow, the deceleration region has a $\sigma_B$ $\gg$ 1 and so the jet is Poynting-flux dominated.

The prompt emission has been suggested to result from magnetic energy dissipation, where the ejecta entrains ordered magnetic fields (\citealt{Lyutikovetal2003} and references therein).
This emission, and the early-time afterglow emission from reverse shocks, may show high levels of linear polarisation in some cases (e.g. \citealt{Steele2009}; \citealt{Mundell2013}; \citealt{Troja2017}). Optical polarisation calibration is well-established, with comparison of GRB measurements and field stars providing additional robustness to detections. More controversial are claims of prompt gamma-ray emission polarisation, with reported measurements spanning the full range from zero to 100\% polarisation, and significant disagreement in the parameter distributions derived with different gamma-ray instruments (e.g., \citealt{Kole2020}).
In a Poynting flux–dominated magnetised jet outflow, the early-time emission is expected to be highly polarised. This is thought to be due to the presence of 
pre-existing magnetic fields, advected from the central source (e.g., see \citealt{Zhang2005} and references therein).
For baryon-dominated jets, the magnetic fields generated locally in shocks are tangled, resulting in unpolarised emission for on-axis jets and low polarisation for edge-on jets \citep{Medvedev1999, Sari1999, Mao2017}. Early-time polarisation measurements of GRB afterglows are therefore crucial for probing the details of the shock physics and for discriminating between different jet models (e.g., \citealt{Mundell2013}). 
A review of past GRB prompt and afterglow polarisation measurements can be found in \cite{Covino2016}.

At late times, in the forward shock regime of the afterglow, the predicted polarisation at optical wavelengths is a strong function of the viewing geometry of the jet (i.e. the opening angle of the jet and our viewing angle with respect to the jet center direction), the internal structure of the jet, and the order and strength of the magnetic field (both within the shock and normal to the shock). Most of these parameters influence the total flux light curve only mildly, but have a large effect on the polarisation as a function of time (see e.g. \citealt{Rossi}), leading to models for the polarisation (amplitude and angle) as a function of time,  which can be tested with high quality data of individual afterglows, as well as the ensemble of measurements of a large number of sources (e.g. \citealt{Wiersema121024}; \citealt{Gill}; \citealt{Stringer}; \citealt{Teboul}).  There are now a few dozen GRBs for which optical polarisation has been detected in their afterglows, and a relatively rich phenomenology is found. Generally speaking, most forward-shock-afterglow polarisation measurements show low levels of linear polarisation (at most a few percent), in many cases with clear signs of variability in both polarisation angle and amplitude. In some high signal-to-noise cases, evidence exists for polarimetric amplitude and angle variability associated with bumps in the optical and X-ray total flux light curve (e.g. \citealt{Greiner}; \citealt{Wiersema091018}). Some afterglows exhibit polarisation signatures supporting the model predictions for homogeneous jets with random fields (e.g. a 90 degree polarisation angle flip,  \citealt{Wiersema121024}), whereas some GRBs more closely follow structured jet models instead (which show no such 90 degree angle change), with possibly an ordered magnetic field component normal to the shock (e.g. \citealt{Gill}; \citealt{Teboul}). In many cases it is not practically possible to obtain high quality polarimetry over a long time period, as most afterglows fade rapidly, and therefore single-epoch measurements of a large number of sources remain important to establish the overall parameter space. The interpretation of polarisation data relies on good multi-wavelength light curves (e.g. to measure the jet collimation angle and the position of the synchrotron break frequencies), and it is therefore important to increase the sample of afterglows with both polarimetric measurements and well-sampled light curves, such as the data set presented in this paper.

A relatively poorly explored polarimetric probe of afterglow physics is multi-wavelength polarimetry, combining near-simultaneous polarisation measurements spanning a wide range of wavelengths, which opens a new window on the afterglow physics (e.g. \citealt{Toma}). Recently, instruments at long wavelengths have become sufficiently sensitive to deliver on this promise for both reverse and forward shock regimes (e.g. \citealt{Laskar}; \citealt{Urata}; \citealt{vanderHorst}). At optical wavelengths, spectro-polarimetry has some diagnostic power in this way as well, particularly if (by chance) any of the synchrotron break frequencies (e.g., the synchrotron cooling frequency) are present near the optical band. Spectro-polarimetry also helps to quantify a key contaminant in afterglow polarimetry studies: the polarisation induced by dust in the GRB host galaxy and in our own Galaxy. Multi-colour polarimetry or spectro-polarimetry are the best ways to quantify this contribution, which is likely to play a non-negligible role in the retrieved polarisation distribution of afterglows and their physical interpretation (see, e.g. \citealt{Lazzati021004};  \citealt{Covino2016}; \citealt{Wiersema121024}; \citealt{Kopac2015}; \citealt{Jordana2020}). To date, the number of afterglows studied with optical spectropolarimetry is limited to just a few cases, e.g. GRB\,020813 (\citealt{Barth2003}), GRB\,021004 (\citealt{Wang}), GRB\,030329 (\citealt{Greiner}) and GRB\,080928 (\citealt{Covino2016}). 
In addition, some spectropolarimetric measurements have been performed {for} the SNe accompanying GRBs (e.g., GRB\,060218, \citealt{Maund}).

Here we report on follow-up optical photometry, spectroscopy and spectro-polarimetry of the optical afterglow of \GRB. Prompt $\gamma$-ray emission was detected on 2019-12-21 20:39:11.42 ($\pm$0.01 s) UT by \textit{AGILE} \citep{Longo2019} and on 2019-12-21 20:39:13 UT by \textit{Swift}/BAT \citep{Laha2019}.\\

\section{GRB 191221B}
GRB 191221B was detected and first reported by the {\it Neil Gehrels Swift Observatory} (henceforth {\it Swift}) Burst Alert Telescope (BAT, \citealt{Barthelmy2005}) on 2019-12-21 at 20:39:13 UT \citep{Laha2019}. {\it Swift} slewed immediately to the burst, repointing its narrow-field instruments, the X-ray telescope (XRT, \citealt{Burrows2005}) and the Ultraviolet and Optical Telescope (UVOT, \citealt{Roming2005}). A bright afterglow was detected both by XRT and UVOT. The BAT light curve shows a complex prompt emission structure with a duration T$_{90}=48\pm16$ s in the 15-350 keV band, and the spectrum can be fit with a simple power-law with index $-1.24\pm0.05$ 

The fluence of GRB 191221B was in the top third of all BAT-detected bursts \citep{Sakamoto2019}. The prompt emission was also reported by {\it AGILE}/MCAL \citep{Longo2019}, {\it AstroSat} CZTI \citep{Gaikwad2019}, {\it Insight}-HXMT/HE \citep{Xue2019}, Konus-{\it Wind} \citep{Frederiks2019} and {\it CALET} \citep{Sugita2019}. Although the {\it AstroSat} CZTI is, in principle, able to observe $\gamma$-ray polarisation, the orientation of the spacecraft was not favorable for a detection of \GRB. The earliest prompt $\gamma$-ray detection was obtained by {\it CALET}, at 20:39:05 UTC, which we adopt as the time of the burst, $T_0$. This was followed by the first reported ground-based detection of a bright (unfiltered magnitude = 10.5 mag) optical transient by MASTER-SAAO at 20:41:35 UT, 150 s after the {\it CALET} burst detection \citep{Lipunov2019a}, although earlier data points were subsequently determined (see next section). The source was so bright that UVOT was able to acquire a grism spectrum, which led to a measurement of the redshift of $z=1.19$ \citep{Kuin2019}, later confirmed and refined by the ESO/VLT X-shooter spectrograph to $z=1.148$ by \cite{Vielfaure2019}, who also reported the presence of an intervening system at $z=0.961$ . The afterglow was also detected in the radio band by ALMA (11.1 hrs after the trigger, \citealt{Laskar2019a}), ATCA (17.5 hrs after the burst, \citealt{Laskar2019b}), and MeerKAT ($30$ days after the trigger, \citealt{Monageng2020}).

\begin{figure*}
	\includegraphics[width=0.85\textwidth]{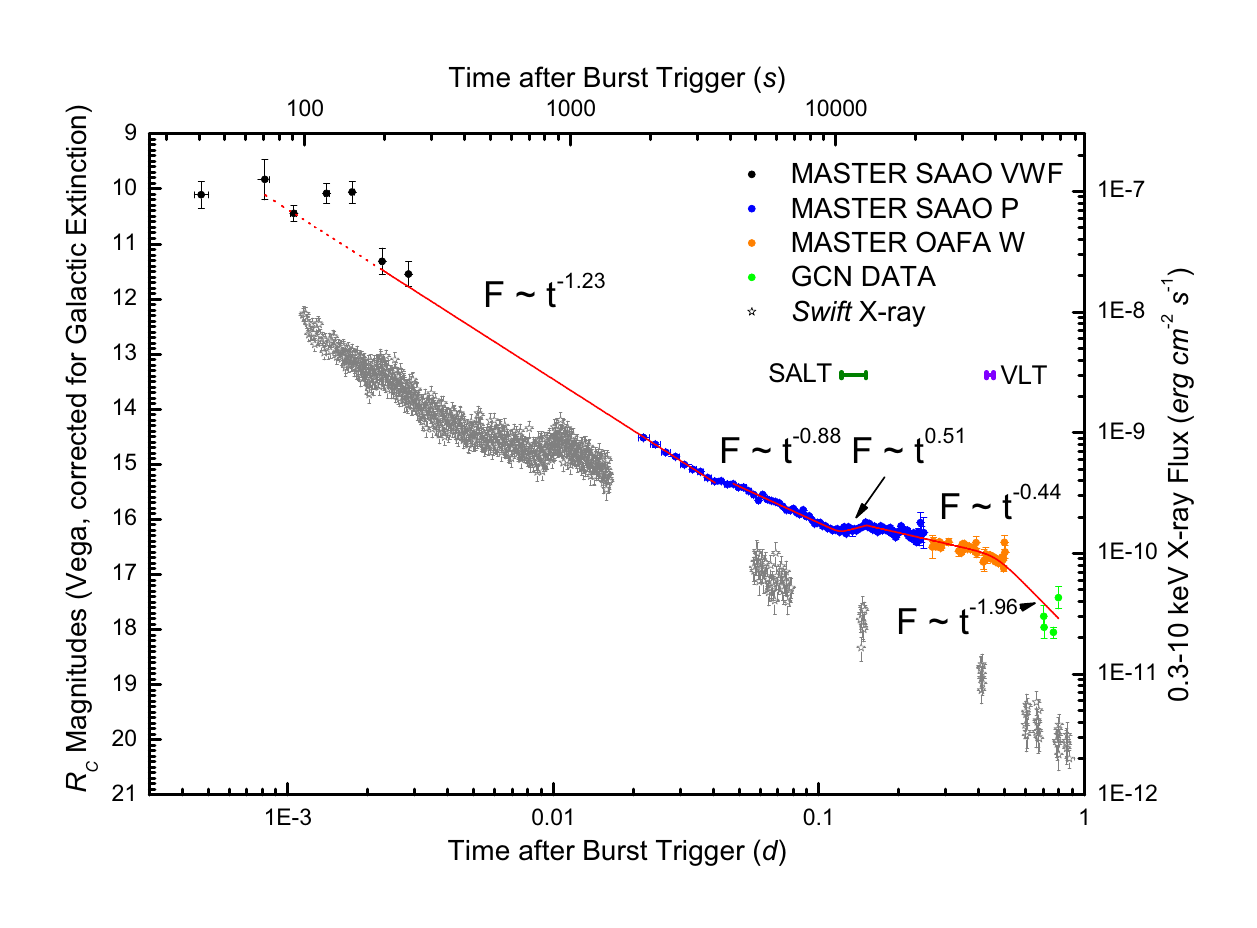}
	\captionsetup{skip=-300pt}
	\caption{Light-curve evolution of \GRB determined by the MASTER-SAAO and MASTER-OAFA facilities (as well as several other ground-based observations, labelled GCN, see text for references), as well as by {\it Swift} XRT in the $0.3-10$ keV range. Time is given in days as well as seconds after the \textit{CALET} burst trigger time, namely $T_0$ = 20:39:05 UT. The spectro-polarimetric coverage by SALT/RSS ($10,472-12,925$ s post-burst) and VLT/FORS2 ($36,906-39,307$ s post burst) is indicated by a green and a purple bar, respectively.}
	\label{fig:MASTER-LC}
\end{figure*}

\section{MASTER photometry of GRB 191221B}
The MASTER Global Robotic Telescope Network \citep{Lipunov2010,Lipunov2019d} began to observe the GRB 191221B error box at 2019-12-21 20:39:43 UT, 38 s post-burst, using the very wide field cameras (VWFC) at MASTER-SAAO, in South Africa \citep{Lipunov2019a}. The VWFC enables wide-field coverage in white light (W) with constant sky imaging every 5~s, which is crucial for GRB prompt detections \citep{Gorbovskoy2010,Kornilov2012,Sadovnichy}. The brightness of the optical afterglow at discovery was $W=10.3$ mag and it remained at this brightness for $\sim$150 s post-burst, thereafter rapidly declining in brightness.  

Observations at MASTER-SAAO, using one of the MASTER-II telescopes (a pair of 0.4 m twin telescopes), started at 2019-12-21 21:09:03 UT ($\sim1798$~s postburst) using a polarizer and clear filter \citep{Lipunov2019b}, although observations were only possible with one of the pair of telescopes due to a CCD camera being non-operational. The position of the optical afterglow was determined by the MASTER auto-detection system \citep{Lipunov2010,Lipunov2019d} from these observations, when \GRB had dimmed to $W=14.4$ mag. The coordinates of the optical counterpart were determined to be RA, Dec. (J2000) $=10^h 19^m 19.24^s, -38^{\circ} 09' 28.7''$ and the optical transient was given the name MASTER OT J101919.24-380928.7 \citep{Lipunov2019c}. MASTER-SAAO observations continued until 21,367 s ($\sim5.93$~h) post-burst, by which time \GRB had faded to $W=16.45$ mag. Observations then began with the 0.4-m MASTER-OAFA 
telescope, in Argentina, 23,017 s ($\sim6.39$~h) post-burst, following the completion of the MASTER-SAAO observations, and continued until 43,324 s ($\sim12$~h) post-burst, at which time the afterglow was at $W=16.77$ mag.

The MASTER clear band magnitude, $W$, is best described by the Gaia $G$ filter. We performed two similar photometric calibration procedures using two different sets of reference stars from the Gaia DR2 catalogue, seven for the VWFC images and nine for the MASTER II telescope images. 
These were used to determine the measurement error (see \citealt{Troja2017} for a more detailed photometric error determination description). After astrometric calibration of each image, we performed standard aperture photometry using ASTROPY/PHOTUTILS \citep{Bradley2016}.  

In \autoref{fig:MASTER-LC} we show the optical light curve evolution of \GRB determined by the MASTER-SAAO and MASTER-OAFA facilities and including subsequent brightness measurements reported in the GCN circulars, as well as by {\it Swift} XRT in the $0.3-10$ keV range, taken from the Burst Analyser \citep{Evans2010}\footnote{https://www.swift.ac.uk/burst\_analyser/00945521/}. 
The light curve of \GRB shows complex breaks in its decline rate, characterised by a general decrease in flux with time, following a sequence of power-laws, $\textit{F} \propto {\it t}^{\alpha}$. The initial decline rate has $\alpha=-1.23\pm0.04$ (measured starting $\approx1,900$ s after the trigger, but a back-extrapolation shows this decay joins with the early very bright emission),
which flattens to $\alpha=-0.88\pm0.02$  at $\sim0.83$~h post-burst. This is followed by a short lived re-brightening, lasting for $\sim0.55$~h and rising with $\alpha=0.51\pm0.14$. The afterglow of \GRB then declined slowly, with $\alpha=-0.44\pm0.01$  until $\sim11$~h post-burst, thereafter breaking and declining more rapidly with $\alpha=-1.96\pm0.14$, where the latter was determined using magnitudes reported in the GCNs \citep{Romanov2019,Gendre2019,Kong2019}. Note the exact value of the latter slope is not well-determined and may change with the addition of further data beyond one day. Details of the power-law slopes are presented in Table \ref{tbl: slopes}.

\begin{table}
\begin{center}
\caption{Power-law slopes of the optical afterglow light curve of \GRB measured at different phases.}
\begin{tabular}{rrc}
T$_{start}$ (s)  & T$_{end}$ (s) & $\alpha$ \\
\hline
1,890 & 3,500 & $-1.23\pm0.04$ \\
3,500 & 10,000 & $-0.88\pm0.02$ \\
10,000 & 12,000 & $\;\;\,0.51\pm0.14$ \\
12,000 & 40,000 & $-0.44\pm0.01$ \\
40,000 & 70,000 & $-1.96\pm0.14$ \\
\hline
\end{tabular}

\label{tbl: slopes}
\end{center}
\end{table}

\begin{figure*}
	\centering
    \includegraphics[width=0.98\textwidth]{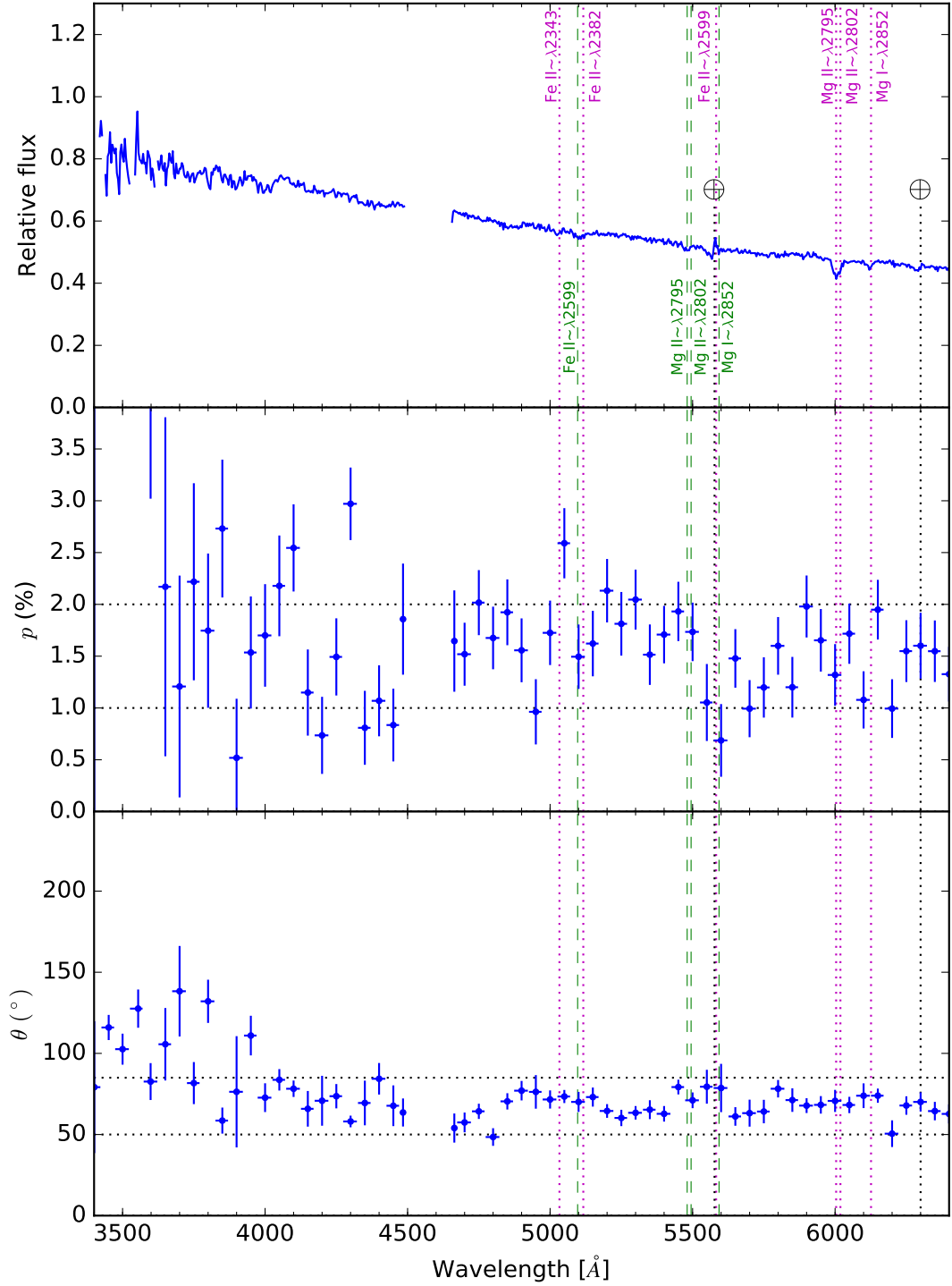}
	\captionsetup{skip=-300pt}
    \caption{SALT/RSS spectropolarimetry of \GRB covering $3400-6300$ \AA, where \textit{p \& $\theta$} were determined after binning the data to 50 \AA\ (see text). Absorption features from the host galaxy ($z=1.15$; magenta dotted lines) and an intervening galaxy ($z=0.96$; green dashed lines) are indicated. Telluric lines are indicated with black dotted lines. There are no data from $\sim$4500--4650 \AA{} due to a chip gap in the CCD mosaic.}
	\label{fig:RSS-specpol-1}
\end{figure*}

\section{Spectropolarimetry}
\subsection{SALT/RSS}
Observations of the optical afterglow of \GRB were obtained with the Southern African Large Telescope (SALT; \citealt{Buckley2006}) using the Robert Stobie Spectrograph (RSS, \citealt{Burgh2003}) in spectropolarimetry mode \citep{Nordsieck2003}. The observations were obtained between 23:34 and 00:15 UTC on 2019 December 21, starting 2 hours 54 min after the GRB alert. The observations were carried out during the re-brightening phase of the light curve.

\begin{figure}
	\centering
	\includegraphics[width=9.4cm,trim={1.8cm 1.5cm 0cm 1.8cm},clip]{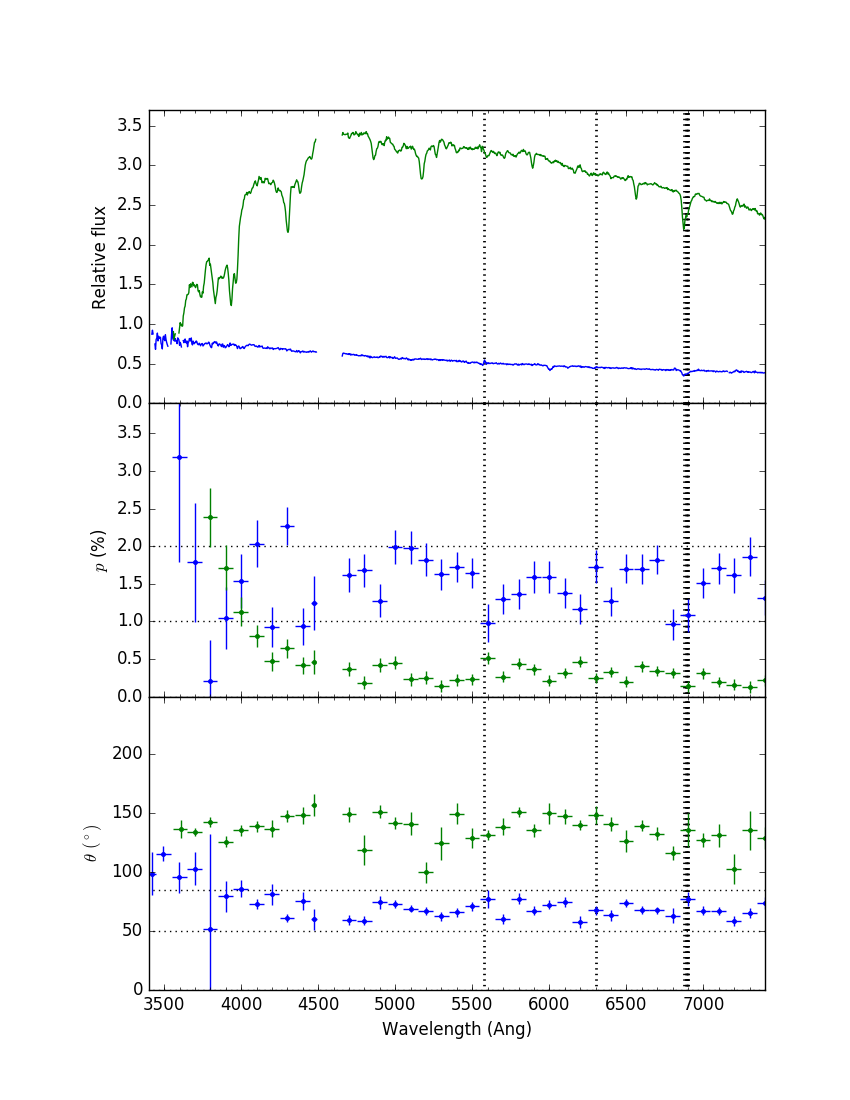}
	\caption{SALT/RSS spectropolarimetry of \GRB (blue) and the nearby bright field star (green), where \textit{p} \& $\theta$ were determined after binning the data to 100 \AA\ (see text). Telluric lines are indicated with thick black dotted lines. There are no data from $\sim$4500--4650 \AA{} due to a chip gap in the CCD mosaic.}
	\label{fig:RSS-specpol-2}
\end{figure}

\begin{figure}
	\centering
	\includegraphics[width=8.8cm,trim={0.8cm 1.5cm 1.0cm 1.8cm},clip]{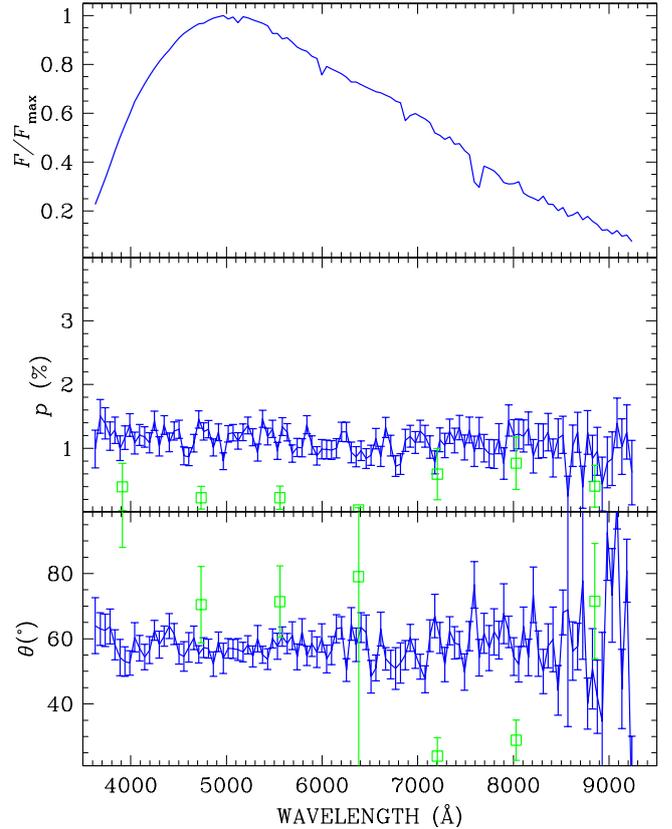}
	\caption{VLT/FORS2 spectropolarimetry of \GRB\ covering $3600-9200$ \AA\ (blue symbols). Data have been rebinned at 50 \AA. Green empty squares show the polarisation of a foreground star, rebinned at 825\,\AA.}
	\label{fig:GRB191221B-FORS}
\end{figure}

Four consecutive exposures of 600~s were obtained at four different orientations of a 1/2 waveplate retarder (0\degr, 22.5\degr, 45\degr\ and 67.5\degr) and the results were analysed to determine the Stokes Q and U parameters, the magnitude of the
linear polarisation, \textit{p}, and the position angle of the E-vector, $\theta$. We used the PG300 transmission grating and a 1\farcs5 wide slit, which gave a wavelength coverage of $3400-8000$ \AA\ at a resolution of $\sim16$ \AA. The spectrograph slit was oriented to a position angle of 45\degr to allow the nearby ($\sim1^\prime$) bright ($B=14.7$, $R=13.2$ mag) reference star, USNO A2 0 0450-11150896, to be measured simultaneously with \GRB. This allows for subtraction of the interstellar polarisation component.

The spectropolarimetry reductions were carried out using an adaptation of the beta version of the polSALT\footnote{\url{https://github.com/saltastro/polsalt}} software\footnote{We used {\sc polSALT} version 20171226 (including {specpolextract\_dev} version 20180524), based on {\sc pySALT} v0.5dev} and the results are shown in \autoref{fig:RSS-specpol-1} and \autoref{fig:RSS-specpol-2} at two resolutions (50 \AA\ and 100 \AA) for the polarisation parameters, the latter figure including the measurements of a nearby field star. We found that \GRB was polarised at an average level of $p=1.5\%$, with a variation of $\pm0.5\%$, and $\theta=65^{\circ}$ with a variation of $\pm10^{\circ}$, over the range $3900-8000$ \AA.

Foreground polarisation due to the ISM was estimated from the nearby ($\sim50\arcsec$ in the SE direction) reference star, USNO A2.0 0450$-$11150896 (Gaia DR2 5444869271098575232), which was also placed on the spectrograph slit. The mean polarisation was $p\sim0.3\%$ and $\theta\sim130\degr$ over the range $4300-7300$ \AA.

\subsection{VLT/FORS2}
Spectropolarimetry of \GRB\ was also obtained using the FORS2 instrument attached at the Cassegrain focus of the Unit 1 (Antu) of the ESO Very Large Telescope. Observations began $\sim10$~h after the burst (from 06:54 UT on 2019 December 22), during the slow decline phase of the afterglow, where  $\alpha=-0.44$. With the 300V grism (with no order separating filter) and a 1\farcs5 slit width. FORS2 observations cover the spectral range from about $3200-9200$ \AA\ with a spectral resolution of $\sim17$ \AA. The observations were performed using the beam-swapping technique, and the total exposure time was 2400~s, equally split in four exposures with the $\lambda/2$ waveplate at position angles 0\degr, 22.5\degr, 45\degr\ and 67.5\degr. Observations were obtained with the E2V blue-optimised CCD mounted on the instrument. Because of the relatively low spectral resolution, fringing at longer wavelengths did not strongly affect the spectrum. Data were reduced using IRAF routines, as described in Sect. 2.3 of \cite{Bagnulo2017}.

The correct alignment of the polarimetric optics was obtained by observing the standard star for linear polarisation, Ve\,6-23 \citep[e.g.][]{Fossati2007} on the same night. In the same slit as the main target, we also observed a foreground star, slightly fainter than the afterglow of \GRB, which showed low polarisation (average $p$ = 0.2\% over 4000--7200 \AA). This reference star was different from the one observed by SALT, being only $\sim5\arcsec$ from \GRB and also considerably fainter. 
The polarisation values of both reference stars are the same, within the uncertainties, indicating a low level of ISM polarisation ($\leqslant$ 0.3\%).

The results are shown in \autoref{fig:GRB191221B-FORS}, where the polarimetric measurements were determined after binning the data to 25 \AA\, per bin. The mean polarisation values determined for \GRB were $p=1.2$\% and $\theta=60$\degr, slightly less than the SALT/RSS values obtained $\sim7$ h earlier.  

\section{Spectral lines}
Significant absorption lines are seen in both the RSS and FORS2 spectra, the strongest located around 6010 \AA\, which was identified as the Mg II 2799/2802 \AA\ doublet by \cite{Vielfaure2019} based on VLT/X-Shooter observations. They concluded that this implied a host galaxy redshift of $z=1.148$. A weaker system of absorption lines, around 5490 \AA, was also seen, corresponding to the same Mg~II doublet at a lower redshift of $z=0.961$, from an intervening system along the line-of-sight.

\begin{figure*}
	\centering
	\includegraphics[width=1\textwidth]{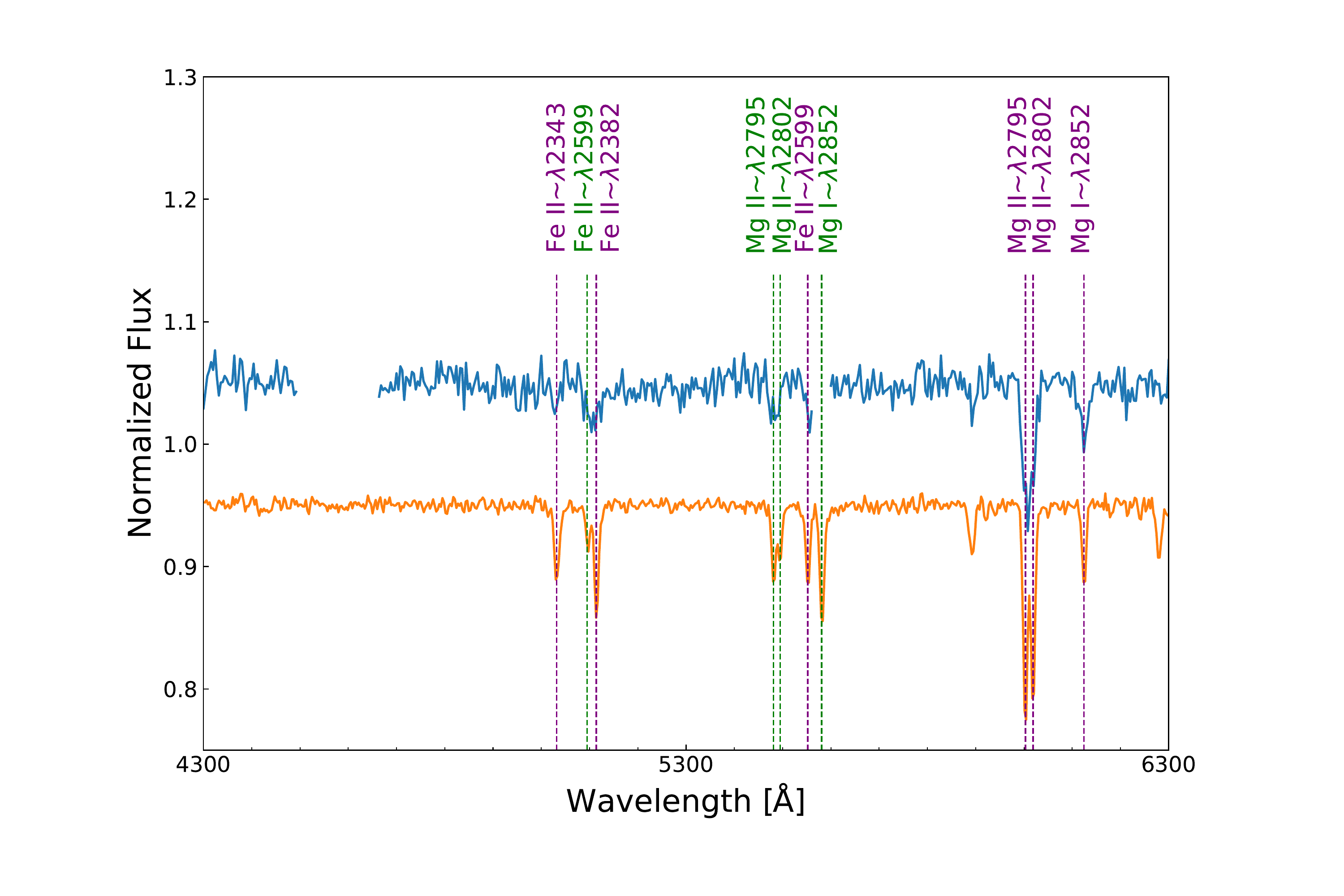}

	\caption{SALT/RSS (blue) and VLT/FORS2 (orange) spectra, normalised and offset by $\pm$ 0.05, respectively. Chip gaps and regions of sky subtraction are omitted from the RSS spectra. Absorption lines from both the host and an intervening galaxy are indicated. The unmarked line at $\approx5900$ \AA\ is Na D ISM absorption. Wavelength labels for the host galaxy lines are in purple while those for the intervening galaxy are in green.} 
	
	\label{fig:spectrum-comps}
\end{figure*}

The \GRB spectra are shown in \autoref{fig:spectrum-comps}, where proposed line identifications are also shown. Line fits were attempted on both spectra and the results are presented for the higher S/N data from FORS2 in Table \ref{tbl: spectrallines}. 
Three close pairs of lines resolved in the FORS2 spectra (Fe II 5096/5114\AA, Mg II 5481/5494\AA\ and Mg II 6002/6018\AA) were unresolved by RSS. This, coupled with the higher noise of the RSS spectra, meant that the higher equivalent width uncertainties precluded making any quantifiable conclusion on any line strength changes between the RSS and FORS2 observations. 

\begin{table*}
\caption{Measurements of spectral lines detected in the optical afterglow of \GRB for the FORS2 observation. Lines corresponding to two redshifts are seen.}
\begin{tabular}{cccccc}
Line ID & Rest Wavelength & Observed Wavelength & FWHM & EW & z  \\ 
& (\AA) & (\AA) &(\AA) &(\AA) &  \\ 
\hline

Fe II & 2343 & 5032.35 $\pm$ 1.44 & 5.44 $\pm$ 1.43 & 0.83 $\pm$ 0.26  &1.148 \\
Fe II & 2599 & 5096.67 $\pm$ 2.19 & 4.70 $\pm$ 2.26 & 0.45 $\pm$ 0.23 & 0.961  \\
Fe II & 2382 & 5114.49 $\pm$ 0.90 & 4.67 $\pm$ 0.96 & 1.10 $\pm$  0.23  & 1.147 \\
Mg II & 2795 & 5481.56 $\pm$ 1.57 & 4.10 $\pm$ 1.67 & 0.67 $\pm$  0.21  &	 0.961 \\
Mg II & 2802 & 5494.59 $\pm$ 2.19 & 3.93 $\pm$ 2.29 & 0.45 $\pm$  0.19 & 0.961  \\
Fe II & 2599 & 5552.48 $\pm$ 1.25 & 4.01 $\pm$  1.25 & 0.62 $\pm$  0.19  & 1.136 \\
Mg I & 2852 & 5581.78 $\pm$ 0.91 & 4.44 $\pm$  0.91 & 1.05 $\pm$  0.22  & 0.957 \\
Mg II & 2795 & 6002.98 $\pm$ 0.37 & 4.56 $\pm$  0.40 & 2.08 $\pm$  0.25 & 1.148 \\
Mg II & 2802 & 6018.71 $\pm$ 0.14 & 4.38 $\pm$  0.43 & 1.83 $\pm$  0.24 & 1.148 \\
Mg I & 2852 & 6124.64 $\pm$ 1.28 & 4.10 $\pm$  1.30 & 0.69 $\pm$  0.22& 1.128 \\
\hline
\end{tabular}
\label{tbl: spectrallines}
\vspace{0.5cm}
\end{table*}

\section{MeerKAT radio observations}
GRB radio afterglows can probe the properties of the jet until very late
times, when the jet essentially becomes non-relativistic. The distribution
of afterglow radio detection times, after trigger, for radio-detected GRBs
peaks between 16 and 32~d, and detections have been made hundreds of days after trigger in some cases \citep{Chandra2012}. 
The typical peak flux density is $\sim100$ $\mu$\,Jy at 8.5\,GHz and $\sim10$~d after
trigger. The radio flux typically declines as {$t^{-1}$} after the peak. 
The radio afterglow of \GRB was detected by ATCA 0.73 days after the GRB, at 5.5, 9.0, 16.7, and 21.2 GHz \citep{Laskar2019b}.
This therefore motivated the attempt to observe \GRB with the MeerKAT radio telescope array \citep{Jonas2009}, in order to detect and monitor any radio emission from this GRB. 

Observations of \GRB with the MeerKAT radio telescope were attempted on 21 January 2020, from 20:26 to 21:26 UTC ($\sim30$\,d after the trigger) and 10 March 2021, from 17:33 to 18:32 UTC ($\sim444$\,d after the trigger), under Director's Discretionary Time \citep{Monageng2020}.  We used J$0408-6545$ as the bandpass and flux calibrator, which was observed for 10~min at the start of the observations. The phase calibrator used was J$1120-2508$, which was observed for 2~min before and after the two $\sim20$ min scans of \GRB in both observations (from 20:41:04.0$-$21:00:55.5 and 21:03:51.5$-$21:23:50.9 UTC on 21 January 2020 and 17:46:19.4$-$18:06:10.9 and 18:09:14.8$-$18:29:06.3 UTC on 10 March 2021, respectively). The observations were performed with 60 antennae and were centered at a frequency of 1.28~GHz with a bandwidth of 856~MHz over 4096 channels. The data were reduced using standard procedures in \textsc{casa} \citep{McMullin2007}. The data were first flagged making use of AOFlagger \citep{McMullin2007}. Thereafter, phase-only and antenna-based delays were corrected for making use of a model based on the primary calibrator. The bandpass correction for the relative system gain over the frequency range of the observation was determined and then complex gains were solved for the primary and secondary calibrators, before scaling the gain corrections for the secondary calibrator from the primary calibrator and applying all the calibrations. Lastly, a small fraction of data were flagged using the \textsc{RFlag} and \textsc{TFCrop} algorithms. Imaging was done using DDFacet \citep{Tasse2018} and self-calibration using the killMS software\footnote{https://github.com/saopicc/killMS} with the \textsc{CohJones} solver. We choose robust $\mathrm{R}=-0.7$ and a cell size of 1\farcs00. For the final direction-independent self-calibrated images, we estimate a rms noise of $\sim17$ $\mu$Jy/beam and $\sim13$ $\mu$Jy/beam for the observations performed on 21 January 2020 and 10 March 2021, respectively, within the vicinity of the source. The dimensions of the synthesised beam are $6.91\arcsec\times 4.44\arcsec$.

\begin{figure*}
	\includegraphics[width=1.0\columnwidth]{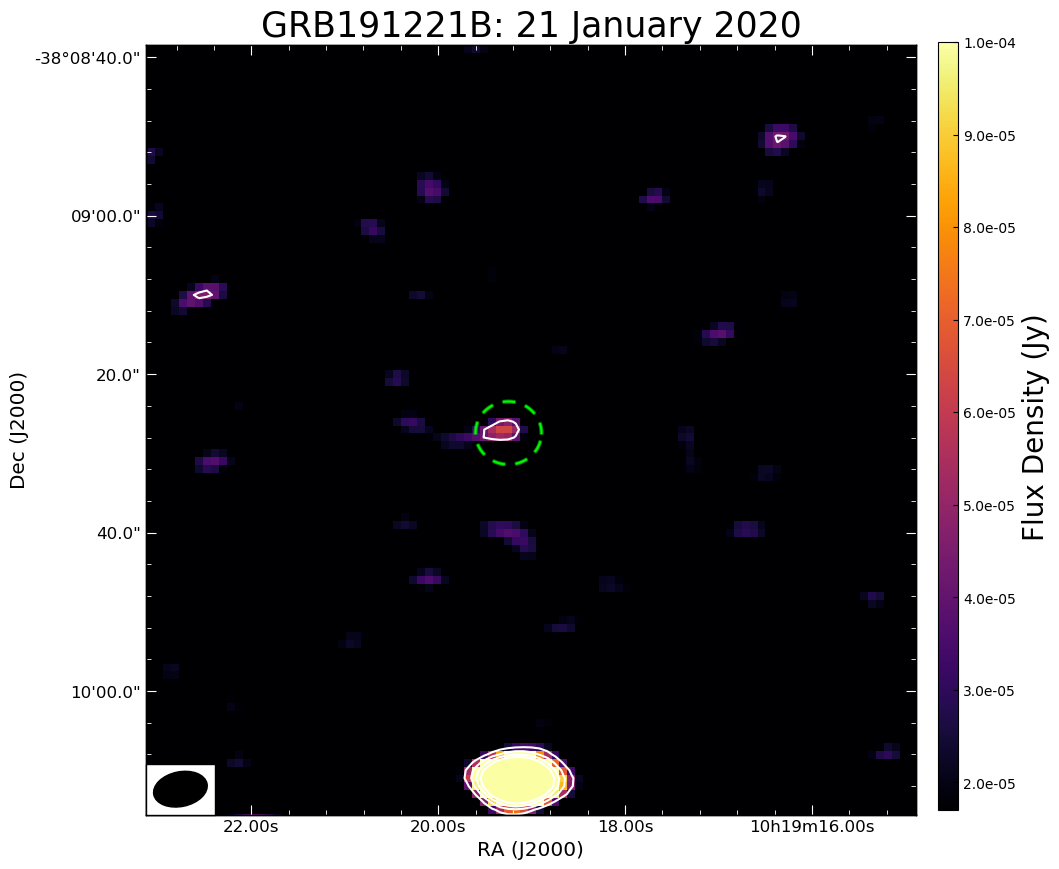}
	\includegraphics[width=1.0\columnwidth]{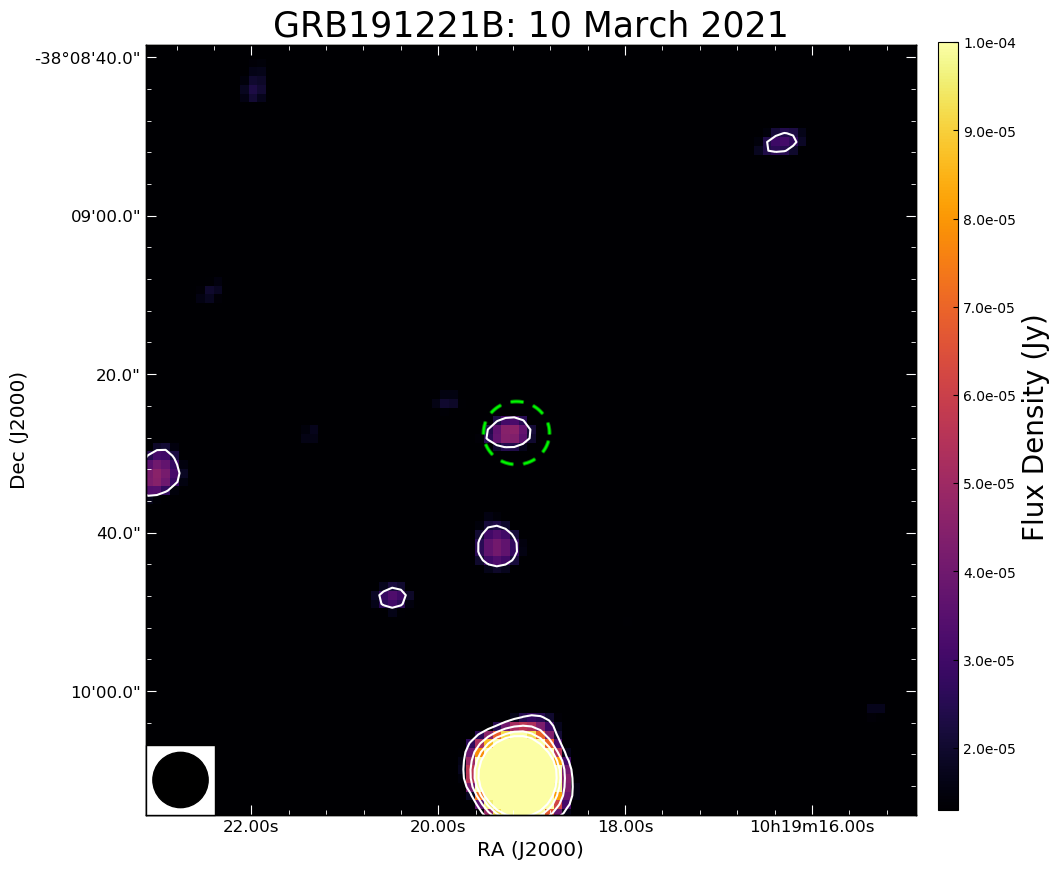}
    \caption{MeerKAT images ($1\farcm5\times1\farcm5$) of the \GRB field, centered at 1.28~GHz and with a bandwidth of 0.86~GHz, for observations performed on the 21 January 2020 (left) and 10 March 2021 (right). The source near the center is coincident with the \GRB position. The green dashed circle is centered at the most accurate position of \GRB, determined from ALMA observations \citep{Laskar2019a}.
	White contours are in multiples of $3\sigma$ and the beam shape is shown in the bottom left corner.}
    \label{fig:GRB-MeerKAT}
\end{figure*}

A source was detected at the nominal \GRB afterglow position, with a peak flux density of $69\pm12$ $\mu$\,Jy/beam ($4.0\sigma$) and $47\pm11$ $\mu$\,Jy/beam ($\sim3.6\sigma$) for the observations performed on 21 January 2020 and 10 March 2021, respectively. We show colour maps of the MeerKAT images of the \GRB field in \autoref{fig:GRB-MeerKAT}, where \sout{the} {\bf{a radio}} source is clearly seen coincident with the optical position.

\section{Models}
The long GRB~191221B had a duration of $T_{90}=13.0\pm1.6$~s \citep{Sugita2019} with multiple pulses during the prompt phase. It was a very 
bright burst, with a 20 keV -- 10 MeV fluence of $(1.0\pm 0.1)\times 10^{-4}$ erg/cm$^2$ \citep{Frederiks2019}. Given the burst was located at a redshift of $z=1.148$ \citep{Vielfaure2019}, GRB~191221B was also rather energetic with $E_{\rm iso} = (3.6\pm 0.1)\times 10^{53}$~erg in the 1 keV -- 10 MeV rest-frame energy range \citep{Frederiks2019}.

Optical observations by MASTER show (see \autoref{fig:MASTER-LC}) a declining flux ($F_\nu \propto t^{-\alpha}$) from $t\gtrsim t_0 + 100$~s, typical of GRB afterglow emission (it is also likely there is an optical flare superposed on the decaying emission, but data coverage is sparse and no deeper conclusions can be drawn). The optical flux decay indices $\alpha_{\rm OPT} =-1.23\pm0.04$ until $t\sim t_0+3.5$~ks and $\alpha_{\rm OPT}=-0.88\pm 0.02$~thereafter, until $t\sim t_0+10$~ks are compatible with synchrotron emission from a forward shock expanding into a constant-density interstellar medium \citep{Sari1999}. In the slow-cooling regime ($\nu_m < \nu < \nu_c$), the forward-shock model predicts $\alpha = 3(p-1)/4 = 0.90-0.98$ for the typically-assumed spectral index $p=2.2-2.3$ for the emitting electrons. 
 
The optical flux decay index is comparable to the {\it Swift} XRT flux (0.3-10~keV) decay index $\alpha_{X} = 1.03^{+0.09}_{-0.08}$ after $t=t_0 + 4.9$~ks \citep{DAvanzo2019}. The X-ray spectral index ($F_{\nu} \propto \nu^{-\beta}$) $\beta_X = 0.86^{+0.07}_{-0.06}$ for the same {\it Swift} XRT flux \citep{DAvanzo2019} is also compatible with the expected value of $\beta = (p-1)/2 = 0.60-0.65$, for $p=2.2-2.3$, from the forward-shock model. Therefore we conclude that both optical and X-ray afterglows of GRB 191221B come from the slow-cooling segment of the synchrotron spectrum. This rules out a reverse-shock origin of the optical afterglow, which predicts $\alpha = (3p+1)/4 =1.90$ for typical values of $p$ \citep{Zhang2003}. 

The relative flattening of the optical light curve after $\sim t_0 + 10$~ks and subsequent decline is expected from the refreshed-shock scenario, where a slower GRB shell ejected later catches up with the decelerating outflow \citep{Panaitescu1998}. 
The light curve after 70~ks, steeply decaying as $t^{-1.96}$ as seen in \autoref{fig:MASTER-LC}, is most-likely due to a jet break, 
which happens when the jet-opening angle $\theta_{\rm jet} \sim 1/\Gamma(t)$, where $\Gamma(t)$ is the bulk Lorentz factor of the jet. A jet break would cause the light curve to decay as $t^{-p}$ after $t\sim t_{\rm jet}$ \citep{Sari1999b}. Assuming the optical flux after 70~ks is post-jet break, the spectral index of the emitting electrons is $p=\alpha = -1.96\pm 0.14$. This is slightly harder than the $2.2-2.3$ values typically inferred from GRB afterglow modeling, but is consistent with generic particle acceleration models within the uncertainties.

The $69\pm 12~\mu$Jy radio flux detected by MeerKAT at 1.28 GHz and $\sim 30$~d after trigger is comparable to the 1.4 GHz flux density measured from other GRBs \citep{Chandra2012}. However, our late-time observation at $\sim444$ days shows that this flux stems mostly or completely from the host galaxy. 
We have estimated what the expected radio flux from star formation in a host galaxy at $z$ = 1.2 or an intervening galaxy at $z$ = 0.96. Using equation 1 presented in \cite{Berger2003}, which is based on the original expression for the observed flux as a function of star formation rate (SFR) derived by \cite{Yun2002}, we calculated the required SFR needed to produce the observed MeerKAT flux. This was 53 M$_{\odot}$ y$^{-1}$, comparable to the SFR presented in \cite{Stanway2014} from radio observations of GRB hosts and consistent with radio observations of star forming region \citep{Murphy2011}.

The flux difference between the two MeerKAT observations is $22\pm16\,\mu$Jy, which may indicate an additional contribution from the radio afterglow, but the difference is not statistically significant. We cannot elaborate on the nature of the radio light curve and derive the power-law decline rate due to scarcity of other radio data points reported for this GRB to date. The flux of the first radio observation by ATCA, obtained 17.5 h after the trigger \citep{Laskar2019b}, is still to be published. 

Synchrotron emission is expected to be highly polarised, although in the context of GRB emission models the expected degree of polarisation is $\lesssim2\%$ for a late afterglow \citep{Covino2016}. 
The reason is that, within the $1/\Gamma(t)$ observable cone there can be a number of magnetic patches, each with a random orientation, thus reducing the degree of polarisation while adding emission incoherently \citep{Gruzinov1999}. This is particularly true for the forward-shock emission, which we believe is the origin of the observed optical emission in \GRB, where the magnetic field is generated from turbulence and the magnetic patches are rather small \citep{Medvedev1999,Mao2017}. Exotic effects such as mixing of photons with axion-like particles can also contribute to polarisation \citep{Mena2011}. The observed level of a few percent linear optical polarisation degree is therefore compatible with this qualitative scenario. An interesting effect is related to the viewing geometry of the afterglow near the time of the jet-break. Around this time, the polarisation components over the area of equal arrival time (an annulus in the case of a homogeneous top-hat jet with a magnetic field that is unordered) no longer sum to zero, and a marked change on the polarisation angle and degree is expected, depending of the viewing angle, the jet opening angle, the jet structure and the order of the magnetic field in the radiating surface \citep{Rossi}. This has been detected in some afterglows (e.g. \citealt{Wiersema121024}) but is not detected in all cases where polarimetry covers times near $t_{\rm jet}$. We see no similar signature in the case of \GRB. Unfortunately, the sparse polarimetric monitoring and the contribution due to dust-induced polarisation in the host galaxy prevents us to draw stronger conclusions. In fact, the various possible configurations model parameters can generate different scenarios for the polarisation time evolution, often with essentially the same flux evolution \citep[e.g.][]{Rossi,Covino2016,Kobayashi2019,Stringer,Shimoda&Toma2020,Teboul,Chengetal2020}.

In addition, if the surrounding medium is dense, we should further consider synchrotron polarisation radiative transfer in the optical band \citep{Mao2018}. The column density determined by the {\it Swift}-XRT observation provides an upper limit of $1.0\times 10^{21}~\rm{cm^{-2}}$, and this corresponds to $A_V=0.56$ mag. The relatively strong absorption leads to the low-degree polarisation of the observed continuum. Alternatively, the low-degree polarisation could be produced by the relativistic electron radiation in the stochastic magnetic field \citep{Medvedev1999, Sari1999, Mao2017}.

It is significant that some absorption lines are clearly detected in the observed spectra of GRB 191221B. Absorption by a patchy dense medium, permeated with a magnetic field, can be strongly amplified by relativistic shocks \citep{Mizuno2014}. When GRB shocks encounter a dense medium, we may consider the possibility of detecting polarisation features in the absorption lines. If the GRB jet is magnetically dominated, the bipolar magnetic field extending along the jet may reach the location of the absorbing material, although the strength of the magnetic field may decrease along the GRB jet. Some material may be ejected by the jet from the GRB central engine (\citealt{Janiuk2014}; see also the recent work on baryon loading in relativistic magnetised shocks by \citealt{Metzger2019}). If particle cooling is effective, the optical photons can be absorbed by the cooled material. In the meanwhile, the magnetic field may have an effect on the material, even at a few parsecs from the GRB central engine.

The theoretical predictions mentioned above might be constrained by spectropolarimetric observations. Polarised radiative transfer of absorption lines was first mentioned by \cite{Unno1956}, where Zeeman splitting produces a triplet structure in a polarised absorption line. The detection of Zeeman split absorption lines is difficult and requires higher resolution spectral observations than in this study. We estimate that a spectral resolution of about R $\sim$ $10^5$ will be necessary if we assume a magnetic field of about $10^3 G$ in the line-forming region \citep{Mao2021}. Although the detection of Zeeman splitting is a hard task, we believe that such a detection in GRB absorption lines in the optical band could be attempted in the future with sufficiently high-resolution spectropolarimetry on 8-m class telescopes, or larger.

\section{Conclusions}

We presented multi-epoch optical observations of the bright, long-duration GRB~191221B with MASTER, SALT \& VLT, as well as radio observations with MeerKAT. We obtained detailed photometric data with MASTER, while spectroplarimetric measurements were performed using data from SALT \& VLT. We detected emission at the position of \GRB with MeerKAT at 1.28 GHz at a flux level of $\sim70\,\mu$Jy and $\sim50\,\mu$Jy at 30 d and 444 d post-burst, respectively, implying this to stem from the host galaxy of \GRB, likely due to star formation. 

The bright ($W$ = 10.3 mag) prompt afterglow was detected with MASTER 38 s post-burst and monitored over $\sim$ 12 h to decline to $W$ = 16.8 mag. The optical light curve after the prompt phase shows a smooth, power-law flux decay, as typically expected from GRB afterglow emission, with several breaks at later time. From the measured decline rates of the optical light curve and the close resemblance between the optical and {\it Swift}-XRT light curves, we conclude that the \GRB optical afterglow is powered by slow-cooling synchrotron emission, ruling out a reverse-shock origin. The flattening and subsequent decline after $\sim10$ ks is attributed to a refreshed-shock scenario, where a faster-moving shell ejected later catches up with the initial decelerated outflow. The steeper decay after 70 ks is likely due to a jet break. We confirm that the inferred spectral index of radiating electrons is typical of the ones expected from the Fermi shock-acceleration process.

Linear polarisation of optical emission from \GRB was first detected by SALT/RSS at $\sim$ 1.5\% some $\sim$3 h post burst, during a period when the brightness had plateaued. Observations with VLT/FORS2 showed little change in polarisation $\sim$ 10 h later, when \GRB was on the decline. Such a low-level polarisation is expected for the late afterglow, when the emission is dominated by the forward shock with a randomly oriented magnetic field configuration. 

\GRB provided an opportunity to observe afterglow polarisation at late time. Our observations show that the degree of polarisation decreases marginally (by $\sim$0.3\%) over a timescale of $\sim$7 h. Future spectro-polarimetric observations from early to late times could probe magnetic field structures in the reverse- and forward-shock regimes, and a transition from the former to the latter.

\section*{Acknowledgments}

Some of the observations presented here were obtained with SALT under programme 2018-2-LSP-001 (PI: DAHB), which is supported by Poland under grant no. MNiSW DIR/WK/2016/07.
Based on observations collected at the European Southern Observatory under ESO programme 0104.D-0600(C).

DAHB and JT acknowledge support through the National Research Foundation (NRF) of South Africa. MB is supported by the South African Research Chairs Initiative (grant no. 64789) of the Department of Science and Innovation and the NRF.\footnote{Any opinion, finding and conclusion or recommendation expressed in this material is that of the authors and the NRF does not accept any liability in this regard.} DMH acknowledges financial support from the NRF and the SAAO. SR is partially supported by NRF with grant No. 111749 (CPRR) and by a University of Johannesburg Research Council grant. DAK acknowledges support from Spanish National Research Project RTI2018-098104-J-I00 (GRBPhot). NPMK acknowledges support by the UK Space Agency. MASTER (equipment) is supported by Lomonosov Moscow State University Development Program. VL, DV are supported by RFBR grant 19-29-11011. CGM acknowledges financial support from Hiroko and Jim Sherwin.

We thank the Director and staff of SARAO for supporting our MeerKAT DDT observation. The MeerKAT telescope is operated by the South African Radio Astronomy Observatory (SARAO), which is a facility of the National Research Foundation, an agency of the Department of Science and Innovation. 

\section*{Data availability}
The data underlying this article will be shared on reasonable request to the corresponding author. Some data (light curves and spectra) are available at the following: https://tinyurl.com/yyd3hty8

\bibliographystyle{mnras}
\bibliography{references}
\newpage
\pagebreak
\onecolumn

\section*{Supplementary Material}
Table \ref{tbl: light-curve} contains the light curve data used to produce Fig. \ref{fig:MASTER-LC}

\begin{longtable}{| p{.15\textwidth} | p{.10\textwidth} |
p{.10\textwidth} |
p{.10\textwidth} |
p{.10\textwidth} |
p{.20\textwidth} |}

\caption{Photometric data of \GRB from MASTER Network}
\\
\hline
JD	&	Time - T$_{0}^{*}$ (sec)	&	Exp Time (sec)		&	Mag	&	Mag	error &	Telescope	\\
\hline
2458839.360913	&	37.886	&	5		&	10.3	&	0.2	&	MASTER-SAAO (VWFC)	\\
2458839.361261	&	67.885	&	5		&	10.0	&	0.4	&	MASTER-SAAO (VWFC) \\
2458839.361492	&	87.885	&	5		&	10.6	&	0.1	&	MASTER-SAAO (VWFC)\\
2458839.361839	&	117.885	&	5		&	10.3	&	0.2	&	MASTER-SAAO (VWFC)	\\
2458839.362187	&	147.887	&	5		&	10.2	&	0.2	&	MASTER-SAAO (VWFC)	\\
2458839.362707	&	192.886	&	5		&	11.5	&	0.2	&	MASTER-SAAO (VWFC)	\\
2458839.363286	&	242.886	&	5		&	11.7	&	0.2	&	MASTER-SAAO (VWFC)	\\
2458839.381286	&	1798.113	&	180		&	14.70	&	0.03	&	MASTER-SAAO	\\
2458839.383617	&	1999.464	&	180		&	14.82	&	0.05	&	MASTER-SAAO	\\
2458839.385944	&	2200.491	&	180		&	14.96	&	0.04	&	MASTER-SAAO	\\
2458839.388273	&	2401.752	&	180		&	15.04	&	0.03	&	MASTER-SAAO	\\
2458839.390599	&	2602.750	&	180		&	15.18	&	0.02	&	MASTER-SAAO	\\
2458839.392927	&	2803.877	&	180		&	15.27	&	0.03	&	MASTER-SAAO	\\
2458839.395258	&	3005.248	&	180		&	15.31	&	0.04	&	MASTER-SAAO	\\
2458839.397602	&	3207.769	&	180		&	15.42	&	0.03	&	MASTER-SAAO	\\
2458839.399930	&	3408.877	&	180		&	15.48	&	0.03	&	MASTER-SAAO	\\
2458839.402257	&	3609.922	&	180		&	15.48	&	0.03	&	MASTER-SAAO	\\
2458839.404584	&	3811.038	&	180		&	15.54	&	0.03	&	MASTER-SAAO	\\
2458839.406912	&	4012.132	&	180		&	15.54	&	0.04	&	MASTER-SAAO	\\
2458839.409241	&	4213.369	&	180		&	15.60	&	0.03	&	MASTER-SAAO	\\
2458839.411570	&	4414.610	&	180		&	15.60	&	0.03	&	MASTER-SAAO	\\
2458839.413916	&	4617.299	&	180		&	15.67	&	0.03	&	MASTER-SAAO	\\
2458839.416244	&	4818.424	&	180		&	15.73	&	0.03	&	MASTER-SAAO	\\
2458839.418575	&	5019.806	&	180		&	15.83	&	0.03	&	MASTER-SAAO	\\
2458839.420901	&	5220.809	&	180		&	15.73	&	0.02	&	MASTER-SAAO	\\
2458839.423232	&	5422.215	&	180		&	15.82	&	0.03	&	MASTER-SAAO	\\
2458839.425559	&	5623.241	&	180		&	15.84	&	0.02	&	MASTER-SAAO	\\
2458839.427886	&	5824.315	&	180		&	15.86	&	0.03	&	MASTER-SAAO	\\
2458839.430216	&	6025.613	&	180		&	15.88	&	0.04	&	MASTER-SAAO	\\
2458839.432546	&	6226.953	&	180		&	15.94	&	0.02	&	MASTER-SAAO	\\
2458839.434874	&	6428.108	&	180		&	16.01	&	0.02	&	MASTER-SAAO	\\
2458839.437212	&	6630.034	&	180		&	16.01	&	0.03	&	MASTER-SAAO	\\
2458839.439543	&	6831.448	&	180		&	15.98	&	0.02	&	MASTER-SAAO	\\
2458839.441873	&	7032.811	&	180		&	16.06	&	0.03	&	MASTER-SAAO	\\
2458839.444201	&	7233.885	&	180		&	16.09	&	0.03	&	MASTER-SAAO	\\
2458839.446527	&	7434.918	&	180		&	16.01	&	0.03	&	MASTER-SAAO	\\
2458839.448858	&	7636.296	&	180		&	16.12	&	0.03	&	MASTER-SAAO	\\
2458839.451190	&	7837.810	&	180		&	16.12	&	0.03	&	MASTER-SAAO	\\
2458839.453547	&	8041.403	&	180		&	16.19	&	0.02	&	MASTER-SAAO	\\
2458839.455877	&	8242.746	&	180		&	16.26	&	0.03	&	MASTER-SAAO	\\
2458839.458206	&	8443.924	&	180		&	16.24	&	0.03	&	MASTER-SAAO	\\
2458839.460536	&	8645.301	&	180		&	16.24	&	0.02	&	MASTER-SAAO	\\
2458839.462864	&	8846.442	&	180		&	16.30	&	0.03	&	MASTER-SAAO	\\
2458839.465194	&	9047.694	&	180		&	16.30	&	0.03	&	MASTER-SAAO	\\
2458839.467524	&	9249.053	&	180		&	16.34	&	0.02	&	MASTER-SAAO	\\
2458839.469854	&	9450.312	&	180		&	16.35	&	0.03	&	MASTER-SAAO	\\
2458839.472181	&	9651.391	&	180		&	16.38	&	0.02	&	MASTER-SAAO	\\
2458839.474507	&	9852.379	&	180		&	16.36	&	0.03	&	MASTER-SAAO	\\
2458839.476839	&	10053.809	&	180		&	16.35	&	0.03	&	MASTER-SAAO	\\
2458839.479165	&	10254.795	&	180		&	16.41	&	0.03	&	MASTER-SAAO	\\
2458839.481496	&	10456.211	&	180		&	16.36	&	0.03	&	MASTER-SAAO	\\
2458839.483824	&	10657.360	&	180		&	16.32	&	0.04	&	MASTER-SAAO	\\
2458839.486178	&	10860.769	&	180		&	16.44	&	0.03	&	MASTER-SAAO	\\
2458839.488508	&	11062.037	&	180		&	16.32	&	0.03	&	MASTER-SAAO	\\
2458839.490834	&	11263.025	&	180		&	16.34	&	0.03	&	MASTER-SAAO	\\
2458839.493181	&	11465.778	&	180		&	16.39	&	0.09	&	MASTER-SAAO	\\
2458839.495526	&	11668.411	&	180		&	16.35	&	0.03	&	MASTER-SAAO	\\
2458839.497852	&	11869.404	&	180		&	16.39	&	0.03	&	MASTER-SAAO	\\
2458839.500183	&	12070.776	&	180		&	16.34	&	0.03	&	MASTER-SAAO	\\
2458839.502515	&	12272.276	&	180		&	16.33	&	0.04	&	MASTER-SAAO	\\
2458839.504846	&	12473.629	&	180		&	16.31	&	0.04	&	MASTER-SAAO	\\
2458839.507173	&	12674.735	&	180		&	16.29	&	0.03	&	MASTER-SAAO	\\
2458839.509501	&	12875.844	&	180		&	16.23	&	0.04	&	MASTER-SAAO	\\
2458839.511968	&	13088.978	&	180		&	16.31	&	0.03	&	MASTER-SAAO	\\
2458839.514297	&	13290.248	&	180		&	16.26	&	0.03	&	MASTER-SAAO	\\
2458839.516624	&	13491.246	&	180		&	16.34	&	0.03	&	MASTER-SAAO	\\
2458839.518951	&	13692.342	&	180		&	16.37	&	0.03	&	MASTER-SAAO	\\
2458839.521277	&	13893.326	&	180		&	16.30	&	0.03	&	MASTER-SAAO	\\
2458839.523604	&	14094.369	&	180		&	16.38	&	0.03	&	MASTER-SAAO	\\
2458839.525935	&	14295.780	&	180		&	16.41	&	0.03	&	MASTER-SAAO	\\
2458839.528261	&	14496.735	&	180		&	16.37	&	0.03	&	MASTER-SAAO	\\
2458839.530590	&	14697.937	&	180		&	16.30	&	0.03	&	MASTER-SAAO	\\
2458839.532921	&	14899.301	&	180		&	16.38	&	0.03	&	MASTER-SAAO	\\
2458839.535252	&	15100.765	&	180		&	16.33	&	0.03	&	MASTER-SAAO	\\
2458839.537603	&	15303.881	&	180		&	16.37	&	0.03	&	MASTER-SAAO	\\
2458839.539935	&	15505.333	&	180		&	16.37	&	0.03	&	MASTER-SAAO	\\
2458839.542266	&	15706.764	&	180		&	16.40	&	0.03	&	MASTER-SAAO	\\
2458839.544595	&	15907.991	&	180		&	16.33	&	0.04	&	MASTER-SAAO	\\
2458839.546929	&	16109.663	&	180		&	16.43	&	0.03	&	MASTER-SAAO	\\
2458839.549258	&	16310.860	&	180		&	16.41	&	0.04	&	MASTER-SAAO	\\
2458839.551589	&	16512.280	&	180		&	16.39	&	0.03	&	MASTER-SAAO	\\
2458839.553921	&	16713.722	&	180		&	16.50	&	0.04	&	MASTER-SAAO	\\
2458839.556252	&	16915.154	&	180		&	16.41	&	0.03	&	MASTER-SAAO	\\
2458839.558586	&	17116.814	&	180		&	16.42	&	0.03	&	MASTER-SAAO	\\
2458839.560912	&	17317.783	&	180		&	16.41	&	0.04	&	MASTER-SAAO	\\
2458839.563238	&	17518.734	&	180		&	16.30	&	0.02	&	MASTER-SAAO	\\
2458839.565568	&	17720.031	&	180		&	16.35	&	0.04	&	MASTER-SAAO	\\
2458839.567894	&	17921.041	&	180		&	16.37	&	0.04	&	MASTER-SAAO	\\
2458839.570242	&	18123.883	&	180		&	16.40	&	0.03	&	MASTER-SAAO	\\
2458839.572569	&	18324.960	&	180		&	16.38	&	0.04	&	MASTER-SAAO	\\
2458839.574911	&	18527.287	&	180		&	16.37	&	0.03	&	MASTER-SAAO	\\
2458839.577239	&	18728.385	&	180		&	16.49	&	0.03	&	MASTER-SAAO	\\
2458839.579798	&	18949.493	&	180		&	16.47	&	0.03	&	MASTER-SAAO	\\
2458839.582128	&	19150.855	&	180		&	16.52	&	0.03	&	MASTER-SAAO	\\
2458839.584460	&	19352.302	&	180		&	16.45	&	0.03	&	MASTER-SAAO	\\
2458839.586789	&	19553.505	&	180		&	16.43	&	0.05	&	MASTER-SAAO	\\
2458839.589120	&	19754.890	&	180		&	16.50	&	0.04	&	MASTER-SAAO	\\
2458839.591451	&	19956.296	&	180		&	16.59	&	0.05	&	MASTER-SAAO	\\
2458839.593776	&	20157.237	&	180		&	16.42	&	0.04	&	MASTER-SAAO	\\
2458839.596104	&	20358.352	&	180		&	16.41	&	0.06	&	MASTER-SAAO	\\
2458839.598439	&	20560.075	&	180		&	16.56	&	0.10	&	MASTER-SAAO	\\
2458839.600775	&	20761.949	&	180		&	16.24	&	0.18	&	MASTER-SAAO	\\
2458839.603106	&	20963.353	&	180		&	16.45	&	0.15	&	MASTER-SAAO	\\
2458839.607783	&	21367.372	&	180		&	16.42	&	0.28	&	MASTER-SAAO	\\
2458839.626882	&	23017.597	&	180		&	16.68	&	0.21	&	MASTER-OAFA	\\
2458839.629443	&	23238.827	&	180		&	16.58	&	0.07	&	MASTER-OAFA	\\
2458839.632009	&	23460.526	&	180		&	16.65	&	0.05	&	MASTER-OAFA	\\
2458839.634558	&	23680.738	&	180		&	16.67	&	0.06	&	MASTER-OAFA	\\
2458839.637156	&	23905.233	&	180		&	16.66	&	0.04	&	MASTER-OAFA	\\
2458839.639707	&	24125.602	&	180		&	16.66	&	0.03	&	MASTER-OAFA	\\
2458839.642252	&	24345.503	&	180		&	16.60	&	0.04	&	MASTER-OAFA	\\
2458839.644827	&	24568.042	&	180		&	16.59	&	0.03	&	MASTER-OAFA	\\
2458839.647407	&	24790.890	&	180		&	16.70	&	0.04	&	MASTER-OAFA	\\
2458839.649958	&	25011.358	&	180		&	16.63	&	0.05	&	MASTER-OAFA	\\
2458839.667416	&	26519.669	&	180		&	16.59	&	0.06	&	MASTER-OAFA	\\
2458839.695850	&	28976.410	&	180		&	16.75	&	0.10	&	MASTER-OAFA	\\
2458839.698928	&	29242.351	&	180		&	16.74	&	0.04	&	MASTER-OAFA	\\
2458839.701497	&	29464.291	&	180		&	16.77	&	0.04	&	MASTER-OAFA	\\
2458839.704050	&	29684.852	&	180		&	16.78	&	0.03	&	MASTER-OAFA	\\
2458839.706631	&	29907.875	&	180		&	16.62	&	0.03	&	MASTER-OAFA	\\
2458839.709197	&	30129.602	&	180		&	16.65	&	0.03	&	MASTER-OAFA	\\
2458839.711809	&	30355.230	&	180		&	16.62	&	0.03	&	MASTER-OAFA	\\
2458839.714386	&	30577.933	&	180		&	16.67	&	0.02	&	MASTER-OAFA	\\
2458839.716950	&	30799.477	&	180		&	16.68	&	0.03	&	MASTER-OAFA	\\
2458839.719521	&	31021.601	&	180		&	16.65	&	0.02	&	MASTER-OAFA	\\
2458839.722131	&	31247.108	&	180		&	16.71	&	0.02	&	MASTER-OAFA	\\
2458839.724690	&	31468.190	&	180		&	16.68	&	0.03	&	MASTER-OAFA	\\
2458839.727253	&	31689.622	&	180		&	16.66	&	0.03	&	MASTER-OAFA	\\
2458839.729816	&	31911.019	&	180		&	16.69	&	0.02	&	MASTER-OAFA	\\
2458839.732363	&	32131.087	&	180		&	16.72	&	0.03	&	MASTER-OAFA	\\
2458839.734936	&	32353.405	&	180		&	16.69	&	0.02	&	MASTER-OAFA	\\
2458839.737489	&	32574.006	&	180		&	16.71	&	0.03	&	MASTER-OAFA	\\
2458839.740042	&	32794.579	&	180		&	16.69	&	0.03	&	MASTER-OAFA	\\
2458839.742633	&	33018.483	&	180		&	16.73	&	0.03	&	MASTER-OAFA	\\
2458839.745241	&	33243.775	&	180		&	16.72	&	0.03	&	MASTER-OAFA	\\
2458839.747802	&	33465.080	&	180		&	16.76	&	0.05	&	MASTER-OAFA	\\
2458839.750351	&	33685.274	&	180		&	16.60	&	0.11	&	MASTER-OAFA	\\
2458839.752973	&	33911.863	&	180		&	16.79	&	0.08	&	MASTER-OAFA	\\
2458839.775450	&	35853.819	&	180		&	16.95	&	0.13	&	MASTER-OAFA	\\
2458839.778486	&	36116.156	&	180		&	16.92	&	0.20	&	MASTER-OAFA	\\
2458839.781529	&	36379.022	&	180		&	16.92	&	0.05	&	MASTER-OAFA	\\
2458839.784622	&	36646.263	&	180		&	16.80	&	0.11	&	MASTER-OAFA	\\
2458839.787212	&	36870.063	&	180		&	16.88	&	0.02	&	MASTER-OAFA	\\
2458839.789787	&	37092.529	&	180		&	16.88	&	0.10	&	MASTER-OAFA	\\
2458839.792396	&	37318.001	&	180		&	16.84	&	0.03	&	MASTER-OAFA	\\
2458839.795055	&	37547.725	&	180		&	16.85	&	0.03	&	MASTER-OAFA	\\
2458839.797663	&	37773.064	&	180		&	16.83	&	0.02	&	MASTER-OAFA	\\
2458839.800223	&	37994.188	&	180		&	16.88	&	0.02	&	MASTER-OAFA	\\
2458839.802759	&	38213.307	&	180		&	16.83	&	0.03	&	MASTER-OAFA	\\
2458839.805312	&	38433.955	&	180		&	16.86	&	0.02	&	MASTER-OAFA	\\
2458839.807851	&	38653.318	&	180		&	16.89	&	0.02	&	MASTER-OAFA	\\
2458839.810401	&	38873.625	&	180		&	16.87	&	0.03	&	MASTER-OAFA	\\
2458839.812995	&	39097.768	&	180		&	16.90	&	0.03	&	MASTER-OAFA	\\
2458839.815571	&	39320.273	&	180		&	16.86	&	0.03	&	MASTER-OAFA	\\
2458839.818115	&	39540.079	&	180		&	16.90	&	0.03	&	MASTER-OAFA	\\
2458839.820691	&	39762.675	&	180		&	16.95	&	0.02	&	MASTER-OAFA	\\
2458839.823239	&	39982.790	&	180		&	16.91	&	0.02	&	MASTER-OAFA	\\
2458839.825809	&	40204.872	&	180		&	16.92	&	0.03	&	MASTER-OAFA	\\
2458839.828356	&	40424.940	&	180		&	16.93	&	0.02	&	MASTER-OAFA	\\
2458839.830899	&	40644.642	&	180		&	16.90	&	0.02	&	MASTER-OAFA	\\
2458839.833450	&	40865.050	&	180		&	16.98	&	0.02	&	MASTER-OAFA	\\
2458839.836132	&	41096.760	&	180		&	16.91	&	0.02	&	MASTER-OAFA	\\
2458839.838682	&	41317.054	&	180		&	16.93	&	0.03	&	MASTER-OAFA	\\
2458839.841276	&	41541.203	&	180		&	16.97	&	0.03	&	MASTER-OAFA	\\
2458839.843828	&	41761.694	&	180		&	16.93	&	0.02	&	MASTER-OAFA	\\
2458839.846500	&	41992.590	&	180		&	16.89	&	0.03	&	MASTER-OAFA	\\
2458839.849069	&	42214.559	&	180		&	16.86	&	0.03	&	MASTER-OAFA	\\
2458839.851630	&	42435.797	&	180		&	16.92	&	0.03	&	MASTER-OAFA	\\
2458839.854205	&	42658.243	&	180		&	16.88	&	0.05	&	MASTER-OAFA	\\
2458839.856776	&	42880.399	&	180		&	17.07	&	0.05	&	MASTER-OAFA	\\
2458839.859305	&	43098.882	&	180		&	16.59	&	0.13	&	MASTER-OAFA	\\
2458839.861916	&	43324.502	&	180		&	16.77	&	0.12	&	MASTER-OAFA \\
\hline
\label{tbl: light-curve}
\end{longtable}
$^*$ Burst time = T$_0$ = JD 2458839.360475 ({\it CALET}, \citealt{Sugita2019})

\bsp    
\label{lastpage}
\end{document}